%% file: main.tex
\pdfoutput=1
\documentclass{article}
\input{src/others/math_commands}

\usepackage[preprint,nonatbib]{neurips_2023}

\usepackage[utf8]{inputenc} % allow utf-8 input
\usepackage[T1]{fontenc}    % use 8-bit T1 fonts     % hyperlinks
\usepackage{url}            % simple URL typesetting
\usepackage{booktabs}       % professional-quality tables
\usepackage{amsfonts}       % blackboard math symbols
\usepackage{nicefrac}       % compact symbols for 1/2, etc.
\usepackage{microtype}      % microtypography
\usepackage{xcolor}         % colors
\usepackage{graphicx}
\usepackage{tabularborder}
\usepackage[ruled,vlined]{algorithm2e}
\usepackage{stfloats}
\usepackage{wrapfig}
\usepackage{hyperref}
\usepackage{adjustbox}
\usepackage{subcaption}
\usepackage{cmap}
\usepackage{microtype}
\hypersetup{
  colorlinks=true,
  linkcolor=blue!80!black,
  urlcolor=blue!80!black,
  citecolor=blue!80!black
}
\microtypesetup{expansion=true, stretch=40}

\usepackage{diagbox}
\usepackage{etoc}
\etocdepthtag.toc{mtchapter}
\etocsettagdepth{mtchapter}{subsection}
\etocsettagdepth{mtappendix}{none}

\usepackage[capitalize,noabbrev]{cleveref}
\usepackage{appendix}
\usepackage{cleveref}
\crefname{section}{Section}{Sections}
\crefname{theorem}{Theorem}{Theorems}
\crefname{lemma}{Lemma}{Lemmas}
\crefname{equation}{Equation}{Equations}
\crefname{proposition}{Proposition}{Propositions}
\crefname{claim}{Claim}{Claims}
\crefname{appendix}{Appendix}{Appendices}
\crefname{algorithm}{Algorithm}{Algorithms}
\crefname{figure}{Figure}{Figs}
\crefname{table}{Table}{Tables}
\crefname{remark}{Remark}{Remarks}
\crefname{definition}{Definition}{Definitions}
\crefname{equation}{Equation}{Equations}
\crefname{corollary}{Corollary}{Corollaries}
\crefalias{section}{appendix}

\definecolor{codegreen}{rgb}{0,0.6,0}
\definecolor{codegray}{rgb}{0.5,0.5,0.5}
\definecolor{codepurple}{rgb}{0.58,0,0.82}
\definecolor{backcolour}{rgb}{0.95,0.95,0.92}
\definecolor{lightgreen}{HTML}{30E1C8}
\definecolor{lightblue}{HTML}{0254D6}
\definecolor{cite_color}{HTML}{114083}
\definecolor{link_color}{RGB}{153, 0,0}  
\definecolor{url_color}{RGB}{153, 102,  0}
\definecolor{emp_color}{RGB}{0,0,255}

\usepackage{amsmath}
\usepackage{amssymb}
\usepackage{mathtools}
\usepackage{amsthm}
\usepackage{amsmath}
\allowdisplaybreaks[4]

\theoremstyle{plain}
\newtheorem{theorem}{Theorem}[section]
\newtheorem{proposition}[theorem]{Proposition}

\theoremstyle{definition}
\newtheorem{definition}[theorem]{Definition}

\theoremstyle{remark}

\usepackage{xspace}
\newcommand{\ours}[0]{SyNDock\xspace}

\title{SyNDock: N Rigid Protein Docking via \\ Learnable Group Synchronization}

\author{Yuanfeng Ji\textsuperscript{\rm 1}, Yatao Bian\textsuperscript{\rm 2*}, Guoji Fu\textsuperscript{\rm 3}, Peilin Zhao\textsuperscript{\rm 2},
\textbf{Ping Luo}\textsuperscript{\rm 1}\thanks{Corresponding authors: \url{pluo@cs.hku.hk}, \url{yatao.bian@gmail.com}}
\\
\textsuperscript{\rm 1} The University of Hong Kong \\
\textsuperscript{\rm 2} Tencent AI Lab 
\textsuperscript{\rm 3} National University of Singapore\\
}

\begin{document}

\maketitle

\input{src/secs/0-abs}

\input{src/secs/1-intro.tex}
\input{src/secs/2-related.tex}
\input{src/secs/3-background.tex}
\input{src/secs/4-syndock.tex}
\input{src/secs/5-exp.tex}

\input{src/secs/6-conclusions.tex}

\bibliography{src/bibs/method.bib}
\bibliographystyle{unsrt}

%%%%%%%%%%%%%%%%%%%%%%%%%%%%%%%%%%%%%%%%%%%%%%%%%%%%%%%%%%%%

\clearpage
\appendix
\begin{center}
	\LARGE \bf {Appendix}
\end{center}
\etocdepthtag.toc{mtappendix}
\etocsettagdepth{mtchapter}{none}
\etocsettagdepth{mtappendix}{subsection}
\tableofcontents

\input{src/secs/7-app}

\end{document}

%% file: src/others/math_commands.tex
\usepackage{amsmath,amsfonts,bm}

\def\eqref#1{equation~\ref{#1}}
\def\1{\bm{1}}

\def\vmu{{\boldsymbol{\mu}}}

\def\vb{{\mathbf{b}}}

\def\vf{{\mathbf{f}}}
\def\vg{{\mathbf{g}}}
\def\vh{{\mathbf{h}}}

\def\vm{{\mathbf{m}}}

\def\vp{{\mathbf{p}}}
\def\vq{{\mathbf{q}}}

\def\vt{{\mathbf{t}}}

\def\vx{{\mathbf{x}}}
\def\vy{{\mathbf{y}}}
\def\vz{{\mathbf{z}}}

\def\mH{{\mathbf{H}}}
\def\mI{{\mathbf{I}}}

\def\mL{{\mathbf{L}}}

\def\mQ{{\mathbf{Q}}}
\def\mR{{\mathbf{R}}}

\def\mT{{\mathbf{T}}}
\def\mU{{\mathbf{U}}}
\def\mV{{\mathbf{V}}}
\def\mW{{\mathbf{W}}}
\def\mX{{\mathbf{X}}}
\def\mY{{\mathbf{Y}}}
\def\mZ{{\mathbf{Z}}}

\def\mSigma{{\mathbf{\Sigma}}}

\DeclareMathAlphabet{\mathsfit}{\encodingdefault}{\sfdefault}{m}{sl}
\SetMathAlphabet{\mathsfit}{bold}{\encodingdefault}{\sfdefault}{bx}{n}

\def\gE{{\mathcal{E}}}

\def\gG{{\mathcal{G}}}

\def\gN{{\mathcal{N}}}

\def\gS{{\mathcal{S}}}

\def\gV{{\mathcal{V}}}

\def\gX{{\mathcal{X}}}
\def\gY{{\mathcal{Y}}}

\def\oS{{\mathrm{S}}}
\def\oT{{\mathrm{T}}}

\newcommand{\R}{\mathbb{R}}

\newcommand{\softmax}{\mathrm{softmax}}

\DeclareMathOperator*{\argmin}{arg\,min}

%% file: src/secs/0-abs.tex
\begin{abstract}
The regulation of various cellular processes heavily relies on the protein complexes within a living cell, necessitating a comprehensive understanding of their three-dimensional structures to elucidate the underlying mechanisms.
While neural docking techniques have exhibited promising outcomes in binary protein docking, the application of advanced neural architectures to multimeric protein docking remains uncertain. 
This study introduces \ours, an automated framework that swiftly assembles precise multimeric complexes within seconds, showcasing performance that can potentially surpass or be on par with recent advanced approaches.
\ours possesses several appealing advantages not present in previous approaches.
Firstly, \ours formulates multimeric protein docking as a problem of learning global transformations to holistically depict the placement of chain units of a complex, enabling a learning-centric solution.
Secondly, \ours proposes a trainable two-step SE(3) algorithm, involving initial pairwise transformation and confidence estimation, followed by global transformation synchronization. This enables effective learning for assembling the complex in a globally consistent manner.
Lastly, extensive experiments conducted on our proposed benchmark dataset demonstrate that \ours outperforms existing docking software in crucial performance metrics, including accuracy and runtime. For instance, it achieves a 4.5\% improvement in performance and a remarkable millionfold acceleration in speed.
\end{abstract}

%% file: src/secs/1-intro.tex
\section{Introduction}
\label{sec:intro}

\begin{wrapfigure}{r}{0.45\textwidth}
    \vspace{-4mm}
    \centering
    \includegraphics[width=0.45\textwidth]{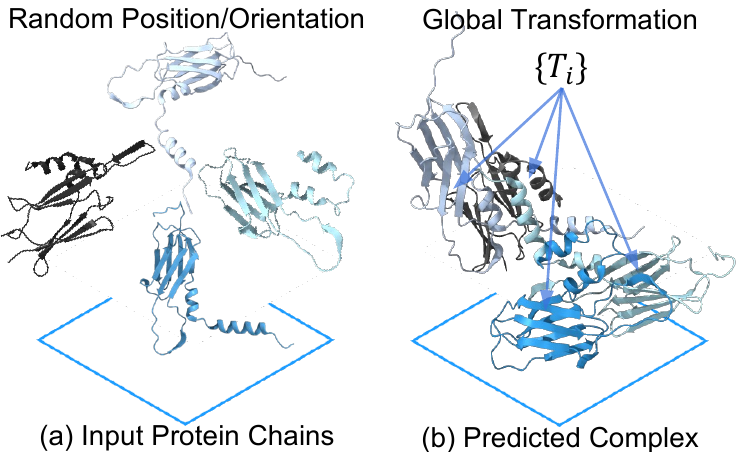}
    \caption{\textbf{Task illustration:} (a) Input multiple proteins and (b) the predicted N-body complex structure. }
    \label{fig:intro:task}
    \vspace{-3.5mm}
\end{wrapfigure}

Protein complexes serve as vital components in numerous biological processes, regulating the expression of essential functions like DNA transcription~\cite{watson1953molecular}, mRNA translation~\cite{nirenberg1961dependence}, and signal transduction~\cite{lemmon2010cell}.
Remarkably, experimental evidence has shown that these functions depend not only on binary complexes but also on the involvement of multimeric complexes.
Structural information about complexes provides critical insight to understand protein function, cellular components, and mechanisms, aiding biomedical research in identifying drug targets, designing therapies, and advancing our knowledge of diseases.
However, the process of detecting the structures of multimeric protein complexes using experimental techniques such as X-ray diffraction~\cite{bragg1913reflection} is often characterized by slowness, high cost, and increased technical challenges compared to solving the structures of single protein chains.
Therefore, the advancement of computational methods for predicting the docking of multimeric protein complexes is of immense importance and offers significant benefits to biomedical research.

Research on the computational prediction of protein complex structures has been conducted for several decades.
Most popular docking software~\cite{lzerd, haddock, attract, zdock, cluspro, rosettadock, swarmdock, hex} is typically limited to the assembly of two protein structures (also known as pairwise docking), and only a few methods~\cite{esquivel2012multi,aderinwale2022rl,ritchie2016spherical,inbar2005prediction} can solve multimeric protein docking, although some of them with many restrictions (i.e., homomeric, symmetric).
Specifically, these algorithms largely follow the steps of coarsely generating a large number (e.g., millions) of possible pairwise docking candidates, followed by combining pairwise solutions using a combinatorial optimization algorithm (e.g., heuristic or genetic algorithm), and further fitting and refining the top-ranked complex structures based on an energy model (e.g., Monte Carlo \cite{multilzerd}). 
However, all of these methods are still computationally expensive and often take between hours and days to predict, with no guarantee of accurately finding complex structures.

\begin{figure*}[t!]
	\centering
	\includegraphics[width=\linewidth]{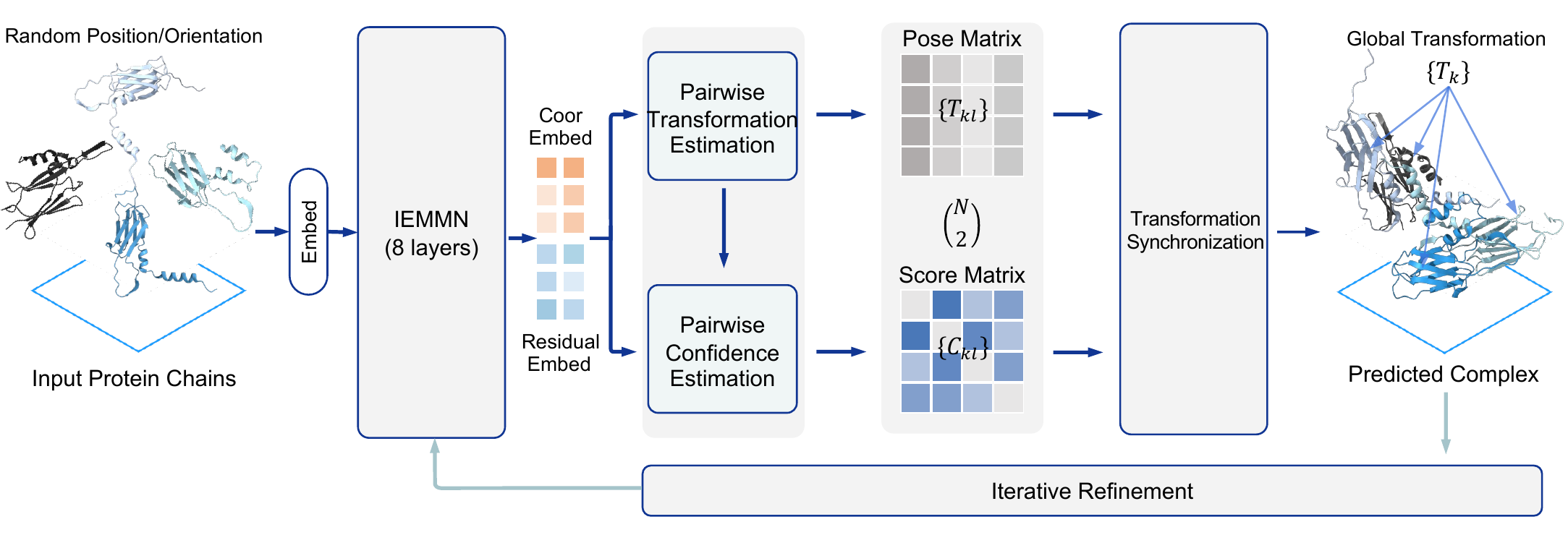}
	\caption{
    \textbf{Overview of \ours,} which comprises three main components: (a) The IEMNN backbone models proteins as graphs and extracts feature embeddings. (b) The pairwise pose estimation module and confidence estimation module utilize the feature embeddings to estimate relative transformations and docking confidence scores between protein pairs. (c) The transformation synchronization module combines these predictions in a learnable manner to recover absolute transformations. The prediction of the multimeric complexes can be easily recovered by applying affine transformations.
    }
	\label{fig:intro:overview}
\end{figure*}

In this paper, we address the problem of \textit{$N$ rigid protein docking}, which refers to predicting the 3D structure of a multimeric complex from the unbound states of individual proteins (\cref{fig:intro:task}).
This problem assumes that the proteins remain rigid during the docking process, without undergoing any deformation, which is a reasonable and widely applicable assumption in various biological contexts.
As an effective solution, we propose \ours, a novel learning-centered approach to multimeric protein docking. Our method formulates the problem as learning global $SE(3)$ transformations for each input protein to recover the exact placement and orientation of protein units within the target complex.
The pipeline of \ours, as shown in \cref{fig:intro:overview}, consists of several steps.
First, a graph-based deep model called the Independent $SE(3)$-Equivariant Multi-graph Matching Network (IEMMN) extracts informative features from the protein chains. 
Next, the relative transformations between pairs of protein chains are estimated, along with the corresponding confidence values.
Finally, a differentiable transformation synchronization module learns to refine the relative transformations to obtain accurate absolute transformations under the "self-consistent" constraint, overcoming potential noise in the input estimations.
These modules are interconnected and trained in an end-to-end fashion and can be iterated to improve performance.

In summary, this paper has three main \textbf{contributions}. 
First, we formulate multimeric protein docking as a problem of learning global transformations to place chain units of a complex, we propose the first end-to-end solution, and it opens a new perspective on multimeric complex docking.
Second, we present a carefully designed, fast, and end-to-end two-step pipeline that enables effective learning to assemble the complex in a globally consistent manner.
Third, we contribute a hetero-multimeric complex dataset curated from the DIPS dataset~\cite{dips}. Extensive experiments show that \ours outperforms recent advanced methods in both accuracy and speed. For example, \ours achieves a 4.5\% performance improvement and a millionfold speedup over Multi-Zerd~\cite{multilzerd} on the tetrameric protein docking task.
These contributions collectively demonstrate the effectiveness and practicality of our approach in advancing the field of multimeric protein docking.

%% file: src/secs/2-related.tex
\section{Related Work}
\paragraph{Protein Structure Prediction.}
Recently, the deep learning based methods AlphaFold2~\cite{alphafold} and Rosettafold~\cite{baek2021accurate} have made a profound impact on the field of structural biology~\cite{naturemethod2021}.
These methods have enabled the prediction of protein structure from primary amino acid sequences and have demonstrated a promising avenue for the tertiary structure determination of proteins.
Inspired by the success of these two methods, there have been several attempts to use pre-trained AlphaFold2 to predict the structure of protein complexes~\cite{ko2021can, tsaban2022harnessing, bryant2022improved} by inserting a linker between two isolated protein sequences.
These methods rely on the slow construction of multiple sequence alignment (MSA) features and are often unable to handle complex docking tasks involving more than two chains.

\paragraph{Protein-Protein Docking.}
Classical experimental methods for determining the structure of protein complexes, including X-ray crystallography~\cite{maveyraud2020protein, dauter2016progress}, nuclear magnetic resonance spectroscopy~(NMR)~\cite{otting2010protein, nitsche2017pseudocontact}, and cryogenic electron microscopy~(Cryo-EM)~\cite{pfab2021deeptracer, costa2017structural, ho2020bottom}.
However, such assays are laborious and expensive, and researchers are trying to solve this problem with computational methods.
As a focal point of activity in computational and structural biology, structure-based docking approaches~\cite{vakser2014protein} are attracting great research interest~\cite{de2010haddock, chen2003zdock, biesiada2011survey, attract2017, weng2019hawkdock, sunny2021fpdock, christoffer2021lzerd, venkatraman2009protein, yan2020hdock, torchala2013swarmdock}. 
They are mainly designed to solve the assembly of paired protein structures, also known as pairwise docking, which typically consists of several main steps: sampling a large number of candidate protein complexes and ranking these candidates with a score function.
Finally, the top candidates are further refined using energy or geometric models~\cite {verburgt2021benchmarking}.
Recently, EquiDock~\cite{ganea2022independent} proposed an end-to-end rigid pairwise protein docking method that is free of the candidate sampling constraint and thus greatly accelerates prediction.
Despite the observed steady progress in the field, methods for assembling three or more chains have received less attention.
The pioneering work~\cite{inbar2003protein,multilzerd} enabled the assembly of multimeric proteins by using optimal combination algorithms from pairwise docked candidates generated by standard binary docking methods.
A series of subsequent papers ~\cite{aderinwale2022rl,ritchie2016spherical} made improvements in various aspects to achieve even better performance.
However, these methods follow the classic paradigm of sampling a large number of complex candidates and further optimizing them using hand-crafted geometric or chemical features to obtain the final structure, resulting in inefficient predictions and unsatisfactory performance.

\paragraph{Transformation Synchronization.}
Given a collection of pairwise estimates, synchronization seeks to synchronize and recover the absolute estimates of latent values that best explain them.
In this paper, we focus on transformation synchronization, which is applicable to both $SO(3)$ and $SE(3)$.
For docking multimeric proteins, one could naively consider only adjacent protein pairs and aggregate the transformations sequentially.
However, this only works if all pairwise estimates are accurate, and an inaccurate/non-existent pairwise alignment will cause the result to fail.
Various approaches have been proposed to more accurately determine the optimal global transformation by considering the additional information contained in the relative transformation, such as the level of confidence.
The method of ~\cite{arrigoni2016spectral,arrigoni2014robust} proposes a closed-form solution for $SO(3)$ and $SE(3)$ synchronization based on the eigendecomposition of a weighted pairwise transformation matrix.
The following approaches~\cite{gojcic2020learning,huang2019learning} build on these ideas and integrate transformation synchronization with a supervised end-to-end multi-view point cloud registration pipeline.
We extend transformation synchronization for the first time to effectively and efficiently predict the structure of multimeric protein complexes.

%% file: src/secs/3-background.tex
\section{Preliminaries and Background}
\paragraph{Problem Statement}
We consider a set of $N$ potentially interacting proteins $\gS = \left\{\gG_k\right\}_{k=1}^N$ as input, which forms a multimeric complex.
Each protein $\gG_k$ consists of $n_k$ residues, represented in their bound (docked) state as 3D point clouds $\mX_k^* \in \R^{3 \times n_k} $, where the position of each residue is given by the coordinate  of its corresponding $\alpha$ carbon atom.
In the unbound state, each undocked protein is arbitrarily rotated and translated in space, resulting in a point cloud $\mX_k \in \R^{3 \times n_k}$ with modified random positions and orientations.
Given arbitrary proteins and their unbound positions $\mX_k$ as input, the rigid protein docking task seeks to calculate the absolute transformations $\mT_k$ such that the result of the affine transformation $\mX_k \otimes \mT_k$ is equal to the desired result $\mX_k^*$.
Intuitively, we represent a transformation $\mT \in SE(3)$ by a $4 \times 4$ matrix. Unless otherwise noted, we denote the rotation and translation components of $\mT$ as $\mR \in SO(3) \subset \R^{3 \times 3 } $ and $\vt \in \R^3$ respectively.
Rather than regressing the global transformations directly, we first estimate the pairwise transformations $\mT_{k,l}$ for each pair of proteins $k$ and $l$, and subsequently exploit transformation synchronization to infer the optimal global transformations $\mT_{k}$ accurately.
Please refer to the \cref{supp:table:notation} in \cref{supp:sec:notapro} for the comprehensive notation of the paper.

\paragraph{Equivariance and Invariance}
To achieve a reliable prediction of the complex structure, we need to satisfy two  constraints: 
(1) \textit{equivariance constraint}: We desire the predicted structure of the proteins to be independent of their initial positions, orientations, and order of input.
(2) \textit{invariance constraint}: We require the final complex structure to be the same after superposition, regardless of the random transformations that are applied.
Formally, we wish to guarantee that:
\begin{definition}[Equivariance]
    Let $\oT_g: \gX \mapsto \gY$ be a set of transformations on $\gX$ for the abstract group $g \in G$. We say that a function $\phi: \gX \mapsto \gY$ is equivariant to $g$ if there exists an equivalent transformation on its output space: $\oS_g: \gY \mapsto \gY$ such that $\phi(\oT_g(\vx)) = \oS_g(\phi(\vx))$.
\end{definition}

\begin{definition}[Rotation Equivariance]
    We say that a function $\phi$ is rotation equivariant if for any orthogonal matrix $\mQ \in \R^{3 \times 3}$, rotating the input $\mX$ resulting in an equivalent rotation of the output $\phi(\mX)$, i.e., $\mQ\phi(\mX) = \phi(\mQ\mX)$. 
\end{definition}

\begin{definition}[Translation Equivariance]
    We say that a function $\phi$ is translation equivariant if for any translation vector $\vg \in \R^3$, translating the input $\mX$ by $\vt$ results in an equivalent translation of the output $\phi(\mX)$, i.e, let $\mX + \vg := (\vx_1 + \vg, \dots, \vx_n + \vg)$, we have $\phi(\mX) + \vg = \phi(\mX + \vg)$. 
\end{definition}

\begin{definition}[$SE(3)$-Equivariance]
    We say that a function $\phi$ is $SE(3)$-equivariant if it is translation equivariant and rotation equivariant.
\end{definition}

\begin{definition}[Permutation Equivariance]
    We say that a function $\phi$ is permutation equivariant if for any column index permutation operator $\sigma$, permuting the input $\mX$ results in the same permutation of the output i.e., $\sigma(\phi(\mX)) = \phi(\sigma(\mX))$.
\end{definition}

\begin{definition}[Invariance]
    Let $\oT_g: \gX \mapsto \gY$ be a set of transformations on $\gX$ for the group $g \in G$. We say that a function $\phi: \gX \mapsto \gY$ is invariant to $g$ if $\phi(\mX) = \phi(\oT_g(\mX))$.
\end{definition}

%% file: src/secs/4-syndock.tex
\section{SyNDock}
\paragraph{Overview}
\label{sec:syn:overview}
The overall structure of SyNDock is outlined in \cref{fig:intro:overview}, which begins by training an encoder $\Phi$ network, represented by a graph neural network (GNN) that is $SE(3)$ equivariant, to extract features from $N$ given protein chains $\gS$.
Then, the pairwise relative transformations $\left\{\mT_{k,l}\right\}_{k=1, k \neq l}^N$ between the $N \choose 2$ pairs of protein chains are estimated along with the corresponding confidence values $\left\{c_{k, l}\right\}$, and the predicted relative parameters are then synchronized to recover the global absolute transformation $\left\{\mT_k\right\}_{k=1}^N$ for each of the individual proteins.
This overall procedure can be further refined through an iterative process, allowing for incremental improvements in the results.
We provide the implementation details in \cref{NagetionAlgo} in \cref{supp:imple} for better clarity.

\begin{wrapfigure}{r}{0.475\textwidth}
    \vspace{-4mm}
    \centering
    \includegraphics[width=0.475\textwidth]{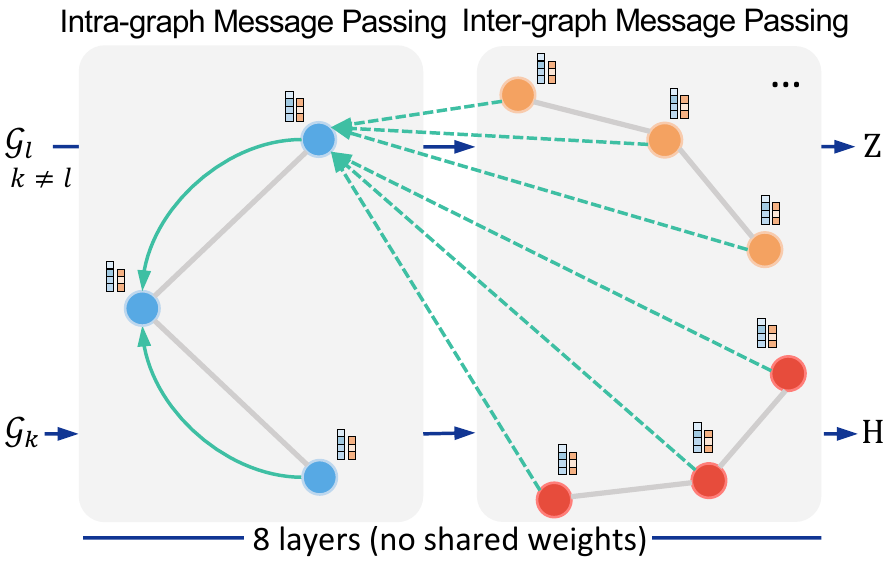}
    \caption{\textbf{IEMMN} message passing}
    \label{fig:intro:encoding}
    \vspace{-3mm}
\end{wrapfigure}

\paragraph{Protein Representation}
\label{src:syn:protein}
Following the work of \cite{jing2020learning,ganea2022independent}, we encode each input protein as a 3D proximity graph, and the graph of protein $k$ is denoted as $\gG_k = (\gV_k, \gE_k)$, where each node $i \in \gV_k$ corresponds to an amino acid with scalar and vector features of $\vh_i \in \R^{f_1}$ and position $\vx_i \in \R^{3}$ corresponding to the Cartesian coordinate of the $\alpha$ carbon atom. 
An edge $(i, j) \in \gE_k$ exists if vertex $j$ is one of the ten most similar neighbors of vertex $i$ based on the Euclidean distance of the coordinates of the two vertices.
Each $(i, j)$ also has edge features that encode both scalar and vector features.
To satisfy the equivariance constraint, we use the $SE(3)$-invariant feature engineering proposed by \cite{ganea2022independent} to construct the additional features $\vf_i \in \R^{f_2}$ and $\vf_{j \rightarrow i} \in \R^{f_3}$ of each node $i \in \cup_{k=1}^N\gV_k$ and each edge $(i, j) \in \cup_{k=1}^N\gE_k$, respectively.
For more details on protein encoding and feature engineering, please refer to \cref{supp:imple}.

% multigraph
As shown in \cref{fig:intro:encoding}, we implement the encoder network $\Phi$ for representation learning by extending the design of  \cite{ganea2022independent}. We name it as  Independent $SE(3)$-Equivariant Multi-Graph Matching Network (IEMMN).
In specific, the $\Phi$ performs node coordinate and feature embedding updating for the input of the protein graphs $\left\{\gG_k\right\}_{k=1}^N$, including the inter- and intra- graph message passing as well as $SE(3)$-equivariant coordinate updates.
The $t$-th layer of the $\Phi$ performs the update of node feature embeddings $\{\vh_i^{(t)}\}_{i \in \cup_k^N\gV_k}$ and node coordinate embeddings $\{\vx_i^{(t)}\}_{i \in \cup_k^N\gV_k}$ as:
\begin{align}
    \vm_{j \rightarrow i}^{(t)} = {} & \phi^{e}\left(\vh_i^{(t)}, \vh_j^{(t)}, \exp(-\left\|\vx_i^{(t)} -\vx_j^{(t)}\right\|^{2} / \sigma ) \vf_{j \rightarrow i}\right), \quad \forall (i, j) \in \cup_{k=1}^{N} \gE_k \label{eq:eq1} \\
    \vmu_{j \rightarrow i}^{(t)} = {} & a_{j \rightarrow i}^{(t)} \mW\vh_{j}^{(t)}, \forall i \in \gV_k, \quad j \in   \cup_{l=1, l\neq k}^N \gV_l  \label{eq:eq2} \\
    \vm_{i}^{(t)} = {} & \frac{1}{|\gN(i)|} \sum_{j \in \gN(i)} \vm_{j \rightarrow i}^{(t)}, \quad \forall i \in \cup_{k=1}^N \gV_k \label{eq:eq3} \\
    \vmu_i^{(t)} = {} & \sum_{j \in \gV_k} \vmu_{j \rightarrow i}^{(t)}, \quad \forall i \in \gV_k, k = 1, \dots, N \label{eq:eq4} \\
    \vx_i^{(t+1)} = {} & \sum_{j \in \gN(i)}\left(\vx_i^{(t)} - \vx_j^{(t)}\right) \phi^{x}\left(\vm_{j \rightarrow i}\right) + \eta\vx_i^{(0)} + (1 - \eta)\vx_i^{(t)}, \quad \forall i \in \cup_{k=1}^N \gV_k \label{eq:eq5} \\
    \vh_i^{(t+1)} = {} & (1-\beta) \cdot \vh_i^{(t)} + \beta \cdot \phi^{h}\left(\vh_i^{(l)}, \vm_i^{(t)}, \vmu_i^{(l)},  \vf_i\right), \quad \forall i \in \mathop{\cup}_{k=1}^{N} \gV_k \label{eq:eq6}
\end{align}

where $\gN(i)$ are the neighbors of node $i ; \phi^x$ is a real-valued (scalar) function; $\mW$ is a learnable matrix; $\phi^h, \phi^e$ are functions outputting a vector $\R^d ; \vf_{j \rightarrow i}$ and $\vf_i$ are the original edge and node features (extracted $SE(3)$-invariantly from the residues); $a_{j \rightarrow i}$ is an attention based coefficient with trainable shallow neural networks $\psi^q$ and $\psi^k$ :
\begin{equation}
    a_{j \rightarrow i}^{(t)} = \frac{\exp\left(\left\langle\psi^q\left(\vh_i^{(t)}\right), \psi^k\left(\vh_j^{(t)}\right)\right\rangle\right)}{\sum_{j^{\prime}} \exp \left(\left\langle\psi^q\left(\vh_i^{(t)}\right), \psi^k\left(\vh_{j^{\prime}}^{(t)}\right)\right\rangle\right)}
\end{equation}
The output of encoder for each protein is then denoted as $\mZ = [\vz_1, \dots, \vz_n] \in \R^{3 \times n}, \mH = [\vh_1^{(T)}, \dots, \vh_n^{(T)}] \in \R^{d \times n}$, where $\vz_i = \vx_i^{(T)}$ for all $i \in \cup_{k=1}^N\gV_k$. $\mZ$ and $\mH$ will be used as the input for subsequent modules.
It is then straightforward to prove the following:

\begin{proposition}\label{thrm:prop1}
    $\vm_{j \rightarrow i}, \vm_i, \boldsymbol{\mu}_{j \rightarrow i}, \boldsymbol{\mu}_i$, and $\vh_i$ in the message passing of IEMMN are invariant to any orthogonal rotation $\mQ \in \R^{3 \times 3}$, any translation vector $\vg \in \R^{3}$. 
\end{proposition}

\begin{proposition}\label{thrm:prop2}
    IEMMN, denoted as $\Phi_\text{IEMMN}$, is rotation equivariant, i.e., for any orthogonal rotation $\mQ \in \R^{3 \times 3}$, at each layer $t$ we have:
$
        \mQ\mX^{(t+1)} = \mQ\Phi_\text{IEMMN}(\mX^{(t)}) = \Phi_\text{IEMMN}(\mQ\mX^{(t)}).
$
\end{proposition}

\begin{proposition}\label{thrm:prop3}
    IEMMN is translation equivariant, i.e., for any translation vector $\vg \in \R^3$, at each layer $t$ we have:
$
        \mX^{(t+1)} + \vg = \Phi_\text{IEMMN}(\mX^{(t)}) + \vg = \Phi_\text{IEMMN}(\mX^{(t)} + \vg).
$
\end{proposition}

\begin{proposition}\label{thrm:prop4}
    IEMMN is permutation equivariant, i.e., for any permutation operator $\sigma$ on the order of $\mX$, at each layer $t$ we have:
$
        \sigma(\mX^{(t+1)}) = \sigma(\Phi_\text{IEMMN}(\mX^{(t)})) = \Phi_\text{IEMMN}(\sigma(\mX^{(t)})).
$
\end{proposition}

Therefore, the node feature embeddings $\mH$ is invariant by \cref{thrm:prop1} and the node coordinate embeddings $\mZ$ is $SE(3)$-equivariant and permutation equivariant by \cref{thrm:prop2,thrm:prop3} and \cref{thrm:prop4}, respectively.

\paragraph{Estimating Pairwise Transformation and Confidence Score}
For each pair of proteins, 
the  goal of pairwise protein docking is to retrieve optimal $\hat{\mR}_{k,l}$ and $\hat{\vt}_{k,l}$:
{\small
\begin{equation}
\hat{\mR}_{k,l}, \hat{\vt}_{k,l} = \underset{\mR_{k, l}, \vt_{k, l}}{\argmin } \sum_{s=1}^{N_{\mathrm{BP}}}\left\|\mR_{k, l} \vp_s + \vt_{k, l} - \varphi\left(\vp_s, \vq_s\right)\right\|^2
\end{equation}}

where $N_{\mathrm{BP}}$ denotes the number of binding pockets between the source and target proteins, and $\phi(\vp_s, \vq_s)$ is a correspondence function that maps the pocket points $\{\vp_s\}_{s=1}^{N_\mathrm{BP}}$ to their corresponding binding points $\{\vq_s\}_{s=1}^{N_\mathrm{BP}}$ in the target protein.
In our setting, the binding pockets of two interacting proteins are matched one by one, so the $\varphi$ can be thought of as an identity function.
In the following, we will show how to adapt the optimal transport loss and Kabsch algorithm in~\cite{ganea2022independent} for pairwise transformation and confidence score prediction.  

\paragraph{Pairwise Transformation Prediction}

In specific, we adopt the multi-head attention mechanism to setup up $M$ key-points queries of each protein of the input pair $\left\{\gG_k, \gG_l\right\}$, 
The set of binding keypoints predictions can be formulated as 
$
    \vy_m^{k} :=\sum_{i \in \gV_k} \alpha_i^{(m)}\vz_i^{k}, \text{ for all } m \in [M]
    % , k \in [N], 
$
where $\vz_i^{k} \in \R^3$ denotes the $i$-th column of the matrix $\mZ_k$, and $\alpha_i^{(m)} = \softmax_i\left(\frac{1}{\sqrt{d}} \mathbf{h}_{i}^{k\top} \mathbf{W}_m^{\prime} \mu\left(\varphi\left(\mathbf{H}_l\right)\right)\right)$ are attention scores, with $\mathbf{W}_m^{\prime} \in \mathbb{R}^{d \times d}$ a parametric matrix (different for each attention head), $\varphi$ a linear layer plus a LeakyReLU non-linearity, and $\mu(\cdot)$ is the mean vector.
We train them to approximate/superimpose the truth binding pockets of the respective protein pair. We drive such a process by assuming a \textit{many-to-many} matching loss (i.e., optimal transport loss), which can be formulated as
$
\mathcal{L}_{\mathrm{OT}}=\min _{\mathbf{T} \in \mathcal{U}(S, K)}\langle\mathbf{T}, \mathbf{C}\rangle
$
\text {, where } $\mathbf{C}_{m, s}=\left\|\mathbf{y}_{k m}-\mathbf{p}_{k s}\right\|^2+\left\|\mathbf{y}_{l m}-\mathbf{p}_{l s}\right\|^2$, 
and $\mathcal{U}(S, M)$  is the set of $S \times M$ transport plans with uniform marginals $\mathbf{p}_{k}$ is the binding points of protein $k$.
Next, given the pocket point prediction ${\mY_k, \mY_l}$, the relative transformation $\mR_{k,l}$ from the source protein to the target protein can easily be calculated by a differentiable ordered point sets registration algorithm (e.g., Kabsch algorithm) as:
$
\mR_{k,l}, \vt_{k,l} = \mathcal{F}_{reg}(\mY_k, \mY_l)
$

\paragraph{Pairwise Confidence Estimation.}
Synchronization is more effective when weights are provided for each of the pairwise estimates.
We weight each pairwise transformation, formulate the estimation of $c_{k, l}$ as a binary classification task, and define the confidence estimation function as
\begin{equation}\label{conf_e}
    c_{k, l} \leftarrow \Phi_{con}\left(\vx_k \otimes \mT_{k,l}, \vx_j, \vh_k, \vh_l\right)
\end{equation}
where the input consists of (i) the coordinates of the transformed source and target proteins, the $\otimes$ is the affine transformation function that aligns the source and target proteins, (ii) the feature/latent vector of proteins.
We use a mean operator to generate features. These features are combined and fed into a confidence estimation network with three fully connected layers (64-64-32-1) and a sigmoid activation function. 
The output of this network is a score $c_{k,l}$ which is a value between 0 and 1.
The corresponding ground truth $c_{k, l}^{\text{GT}}$ is obtained directly from the training data.
To avoid redundant computations, we only estimate pairwise pose predictions and scores for pairs $(k, l)$ where $k < l$. For pairs $k > l$, we set $c_{l,k} = c_{k,l}$, $\mR_{l,k} = \mR_{k,l}^{-1}$, and $\vt_{l,k} = \mR_{k,l} \cdot \vt_{k,l}$.
The training of the pairwise protein docking module is directly supervised by the combination of the relative transformation loss $\mathcal{L}_{\mathrm{REL}}$ and the confidence loss function $\mathcal{L}_{\mathrm{CON}}$, where the $\mathcal{L}_{\mathrm{CON}}$ denotes the binary cross entropy loss, and  
{\small
\begin{equation}
    \mathcal{L}_{\mathrm{REL}} = \sum_{(k, l) \in \gS}\left(\left\|\mR_{k, l} - \mR_{k, l}^{\text{GT}}\right\|_2 + \left\|\vt_{k, l} - \vt_{k, l}^{\text{GT}}\right\|_2\right).
\end{equation}}

\paragraph{SE(3) Transformation Synchronization}
Given the estimated ${N \choose 2}$ relative protein-to-protein transformations, we aim to find the $N$ global absolute transformations $\left\{\mT_{k}\right\}$ that best explain them.
Previous method~\cite{arrigoni2016spectral,arrigoni2014robust} proposed a closed-form solution to $SE(3)$ synchronization using spectral decomposition. This idea was further developed in end-to-end learning pipelines~\cite{gojcic2020learning,huang2019learning}. The approach is based on the construction of a block matrix of pairwise transformations, where the block $(k, l)$ represents the transformation between instance $k$ and $l$. The key insight in this line of research is that the absolute transformations can be extracted from the pairwise transformation matrix by means of eigendecomposition.
The global transformation parameters can be determined either through joint calculation (also known as transformation synchronization) or by breaking the problem into two parts: rotation synchronization and translation synchronization.
In this paper, we opt for the latter approach, dividing the problem into rotational and translational synchronization, which admits a differentiable closed-form solution under the spectral relation.

\paragraph{Rotation Synchronization.} Our approach to rotation synchronization is based on a Laplacian rotation synchronization formulation that has been proposed previously in the literature~\cite{gojcic2020learning}:
$
% \label{sync_r}
    \mR_k^* = \argmin_{\mR_k \in SO(3)}\sum_{(k,l) \in \gE}c_{k,l}\|\mR_{k,l} - \mR_k\mR_l^\top\|_F^2.
$
 More precisely, consider a symmetric matrix $\mathbf{L} \in$ $\R^{3N \times 3N}$, which resembles a block Laplacian matrix, defined as:

\begin{equation*}
    \mL = \begin{bmatrix}
    \mI_3 \sum_i c_{k, 1} & -c_{1,2}\mR_{1,2} & \cdots & -c_{1, N}\mR_{1, N} \\
    -c_{2,1}\mR_{2,1} & \mI_3\sum_ic_{k, 2} & \cdots & -c_{2, N}\mR_{2, N} \\ 
    \vdots & \vdots & \ddots & \vdots \\ -c_{N, 1}\mR_{N, 1} & -c_{N, 2}\mR_{N, 2} & \cdots & \mI_3\sum_i c_{k, N} \\ \end{bmatrix}.
\end{equation*}

where the $c_{k,l}$ represents the pairwise confidence score. $\mR_{k,l}$ is the estimated relative rotation transformation, and the $N$ is denoted as the number of input protein chains.
Consequently, we collect the three eigenvectors  $\mU = \left(\mU_1^\top, \cdots, \mU_N^\top\right)^\top \in \R^{3N \times 3}$ that correspond to the three smallest eigenvalues of $\mL$.
To avoid reflections, we  choose the sign of each eigenvector such that $\sum_{i=1}^N \operatorname{det}\left(\mU_k\right) > 0$.  
The least squares estimates of the global rotation matrices $\mR_i$ are then given, under relaxed orthogonality and determinant constraints,
by first performing singular value decomposition (SVD) on each $\mU_k = \mV_k\mSigma_k\mW_k^\top$
and then output the corresponding absolute rotation estimate as:
\begin{equation}
    \mR_k = \mV_k\mW_k^\top, \quad \text{ for all } k = 1, 2, \dots, N.
\end{equation}

\paragraph{Translation Synchronization.}
We can retrieve global translation vectors $\left\{\mathbf{t}_k\right\}$ that minimize the following least squares problem:
\begin{equation}\label{sync_t}
    \vt_k = \underset{\vt_k}{\argmin} \sum_{(k, l) \in \gE} c_{k, l}\left\|\hat{\mR}_{k, l} \vt_k + \hat{\vt}_{k, l} - \vt_l\right\|^2
\end{equation}
The closed-form solution is easy to get:
$
    \vt = \mL^+\vb,
$
where $\vt^* = [\vt_1^{*\top}, \dots, \vt_{N}^{*\top}]^\top \in \R^{3N}$ and $\vb = [\vb_1^{*\top}, \dots, \vb_{N}^{*\top}]^\top \in \R^{3N}$ with
$
    \vb_k := -\sum_{l \in \gN(k)}c_{k,l}\mR_{k,l}^\top\vt_{k,l}.
$

\paragraph{Iterative Refinement of $N$-body Docking.}
The above formulation allows an implementation in an iterative scheme, with the ability to refine the predictions step by step, starting from a coarse level.
To this end, we can start each subsequent iteration by using the synchronized estimated parameters $\mT_k$ from the previous iteration to fit each protein, setting $\mX_k^* = \mX_k \otimes \mT_k$.
In this paper, we iterate the refinement 4 times for each sample and calculate the loss with the final prediction.

\paragraph{End-to-End Training Algorithm}
\label{sec:syn:e2e}
We supervise the network training using the following supervisions:
$\mathcal{L}_{\mathrm{REL}}$, $\mathcal{L}_{\mathrm{CON}}$, $\mathcal{L}_{\mathrm{OT}}$ as well as the global \textit{Synchronization Loss}:
$
\mathcal{L}_{\mathrm{SYNC}} = \frac{1}{N} \sum_{k=1}^{N}\left\|\mX_k \otimes \mT_k - \mX_k^{\text{GT}} \right\|_2
$
where $\otimes$ is the affine transformation. 
The total loss functions can thus be defined as
\begin{equation}
    \mathcal{L}_{\mathrm{OVERALL}} = \mathcal{L}_{\mathrm{OT}} + \mathcal{L}_{\mathrm{REL}} + \mathcal{L}_{\mathrm{CON}} + \mathcal{L}_{\mathrm{SYNC}}.
\end{equation}

\begin{figure*}[t!]
    \centering
    \includegraphics[width=\linewidth]{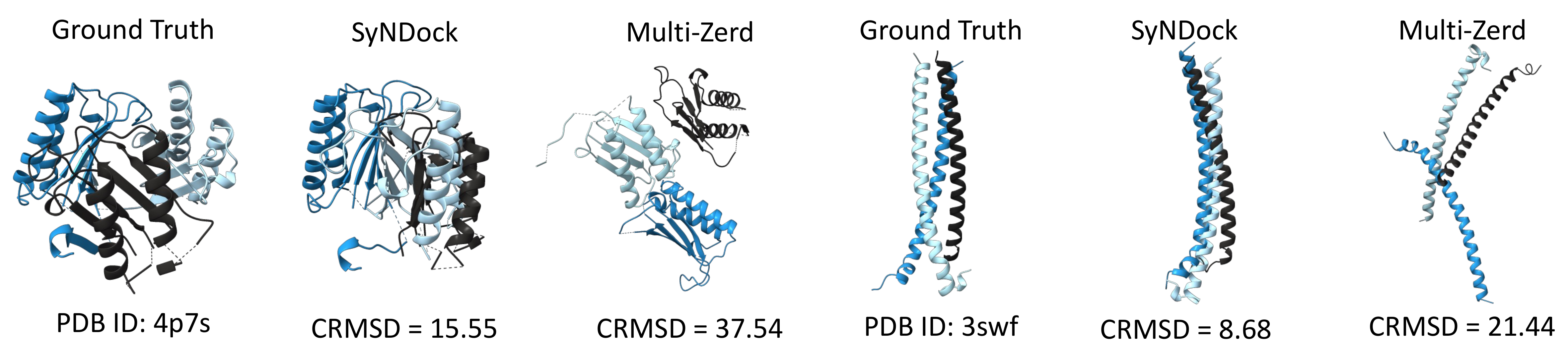}
    \caption{\textbf{Complex Prediction Visualization.} Qualitative comparisons between our approach and baseline approaches on the trimeric protein samples.}
    \label{fig:exp:vis}
\end{figure*}

%% file: src/secs/5-exp.tex
\section{Experiments}

\setlength{\tabcolsep}{8pt}
\renewcommand{\arraystretch}{0.85}
\begin{wraptable}{r}{4.95cm}
	\small
    \centering
    \vspace{-5mm}
    \caption{\textbf{Curated dataset statistics.}}
    \begin{tabular}{clll}
    \toprule
    N-Body & Train & Valid & Test \\ \hline
    2 & 30,299 & 748 & 887 \\
    3 & 19,664 & 441 & 593 \\
    4 & 16,415 & 303 & 429 \\
    5 & 14,105 & 251 & 257 \\
    6 & 10,689 & 281 & 114 \\
    7 & 5,881 & 220 & 30 \\
    8 & 2,143 & 90 & 5 \\
    9 & 545 & 20 & 0 \\
    10 & 53 & 2 & 0 \\
    \hline
    \end{tabular}
    \label{tab:exp:dataset}
\vspace{-0.35cm}
\end{wraptable}

\paragraph{Dataset. }
We evaluated our method on the DIPS dataset \cite{townshend2019end}, which is currently the largest protein complex structure dataset mined from the Protein Data Bank and tailored to the rigid body docking assumption.
We adopted the same split as defined in EquiDock \cite{ganea2022independent}, where the train/validation/test splits are based on the protein family.
After excluding a few structures with more than 10K atoms, we get 14,225 PDB for the training set, 368 for the validation set, and 393 for the test.
To generate subsets of N-body protein complexes, we extracted connected subgraphs with N degrees and removed complexes with more than ten chains.
The statistics of the curated dataset is in \cref{tab:exp:dataset}.
We also performed a binary docking evaluation in \cref{tab:exp:binary}, using the default settings provided by \cite{ganea2022independent}, which included 100 randomly selected test pairs from the DIPS dataset.
We refer the readers to \cref{supp:imple} for more detailed information on the mentioned datasets as well as curation implementation.

\paragraph{Evaluation Metrics.}
We use two metrics to measure the quality of the docking predictions: Complex Root Mean Square Deviation (C-RMSD) and Interface Root Mean Square Deviation (I-RMSD). 
These two metrics are defined below: Given the ground truth $\mathbf{P}^* \in \mathbb{R}^{3 \times\sum_{k=1}^{N}n_k}$ and predicted complex structures $\mathbf{P} \in \mathbb{R}^{3 \times\sum_{k=1}^{N}n_k}$, we first superimpose them using the rigid registration algorithm (i.e., Kabsch) and then compute the C-RMSD as $\sqrt{\frac{1}{\sum_{k=1}^{N}n_k}\left\|\mathbf{P}^*-\mathbf{P}\right\|_F^2}$.
The I-RMSD is calculated similarly, but only takes the coordinates of the residues that are within 8 angstroms of the residues in the other protein.
For a fair comparison between the baselines, we use only the $\alpha$ carbon coordinates for the calculation of both metrics.
\setlength{\tabcolsep}{4pt}
\renewcommand{\arraystretch}{1}
\begin{table*}[t!]
\centering
\small
\caption{\textbf{Multimeric Complex Prediction Result}. Comparison between Multi Zerd~\cite{bregier2021deep} and \ours in two performance metrics in the two curated subsets (N=3,4). All values are smaller the better. We can see that \ours can significantly outperform Multi-Zerd in terms of docking accuracy as well as having a clear advantage in the identification of docking sites.}
\begin{tabular}{lcccccccccccc} \toprule
& \multicolumn{6}{c}{Trimer (N=3)} & \multicolumn{6}{c}{ Tetramer (N=4) } \\
& \multicolumn{3}{c}{ Complex-RMSD $\downarrow$} & \multicolumn{3}{c}{ Interface-RMSD $\downarrow$} & \multicolumn{3}{c}{ Complex-RMSD $\downarrow$} & \multicolumn{3}{c}{ Interface-RMSD $\downarrow$} \\
\hline  
Methods & Median & Mean & Std & Median & Mean & Std & Median & Mean & Std & Median & Mean & Std \\ \hline
Multi-Zerd & 22.17 & 24.06 & 6.08 &23.16 &24.44 & 6.95 &28.51 &29.02 &5.83 &26.75 & 29.23 &8.71\\
\ours & 21.02 & 22.01 & 7.69 &17.80 &18.35 &6.34 & 23.92 & 24.56 &7.12 & 19.75 & 20.37 & 5.85  \\
\bottomrule
\end{tabular}
\label{tab:exp:multi}
\end{table*}

\paragraph{Experimental Configurations.}
We train our model using the AdamW optimizer with a learning rate of $10^{-4}$ and a weight decay of $10^{-3}$. We train for 50 epochs with a batch size of 6 using a fixed learning rate schedule.
Model selection is based on the performance of the validation set. The best validation model is then tested on the test set.
Unless otherwise noted, we train and test separately on different N-body data subsets.
For data augmentation, we randomly change the permutation of inputs during training and testing, and we also randomly rotate and translate all inputs in the space.
Our implementation is based on the PyTorch toolkit~\cite{paszke2017automatic}, with extensive use of the ROMA~\cite{bregier2021deep} package. We will make our code available to the public.
\paragraph{Multimeric Protein Docking.}
We compare \ours to the widely used multimeric protein docking method Multi-Zerd \cite{esquivel2012multi}. 
Note that there are very few open-source, user-friendly multimeric docking programs available for comparison, and some of them (CombDock\footnote{\url{http://bioinfo3d.cs.tau.ac.il/CombDock/download/}}~\cite{inbar2005prediction}, RL-Multi-LzerD\footnote{\url{https://github.com/kiharalab/RL-MLZerD}}~\cite{aderinwale2022rl}) always crash.
For different subsets of N bodies, we randomly select a certain number of test samples from the test set for comparison (60 samples for trimer, 30 samples for tetramer). 
From \cref{tab:exp:multi} we see that \ours significantly outperforms Multi-Zerd, especially in the tetramer docking subset ($N=4$), where the former increases the C-RMSD by up to 4.5\% compared to the latter.
This observation holds true for the I-RMSD metrics as well, demonstrating that our method also has a clear advantage in binding site identification.
Furthermore, our method exhibits a speed improvement of several orders of magnitude compared to the baseline method (see \cref{fig:exp:speed}).
\paragraph{Binary Protein Docking.}
SyNDock can be scaled down to predict binary protein docking.
We compare \ours to the popular binary protein docking baselines: Attract~\cite{attract}, HDock~\cite{yan2020hdock}, ClusPro~\cite{cluspro}, PatchDock~\cite{mashiach2010integrated}, and the recently proposed EquiDock~\cite{ganea2022independent}.
As shown in \cref{tab:exp:binary}, \ours is competitive and often outperforms the baselines, demonstrating that our method also has a general capability.

\setlength{\tabcolsep}{1.2pt}
\renewcommand{\arraystretch}{1.1}
\begin{minipage}[t]{0.49\textwidth}
    \footnotesize
    \centering
    \makeatletter\def\@captype{table}\makeatother\caption{\textbf{Binary Complex Prediction Results}. Comparison of different binary docking methods.}
    \begin{tabular}{lcccccccccccc} \toprule
    & \multicolumn{3}{c}{ Complex-RMSD } & \multicolumn{3}{c}{ Interface-RMSD } \\
    \hline Methods & Median & Mean & Std & Median & Mean & Std  \\ \hline
    Attract & 17.17 & 14.93 & 10.39 & 12.41 & 14.02 & 11.81  \\
    HDock & 6.23 & 10.77 & 11.39 & 3.90 & 8.88 & 10.95 \\
    ClusPro & 15.76 & 14.47 & 10.24 & 12.54 & 13.62 & 11.11 \\
    PatchDock & 15.24 & 13.58 & 10.30 & 11.44 & 12.15 & 10.50 \\
    EquiDock & 13.29 & 14.52 & 7.13 & 10.18 & 11.92 & 7.01 \\ \hline
    \ours & 12.95 & 14.27 & 6.85 &9.97 & 11.93 & 5.84\\
    \bottomrule
    \end{tabular}
    \label{tab:exp:binary}
\end{minipage}\hfill
\begin{minipage}[t]{0.49\textwidth}
    \footnotesize
    \centering
    \makeatletter\def\@captype{figure}\makeatother\caption{\textbf{Inference time distribution.} Both methods are tested on the same hardware.}
    \includegraphics[width=0.8\linewidth]{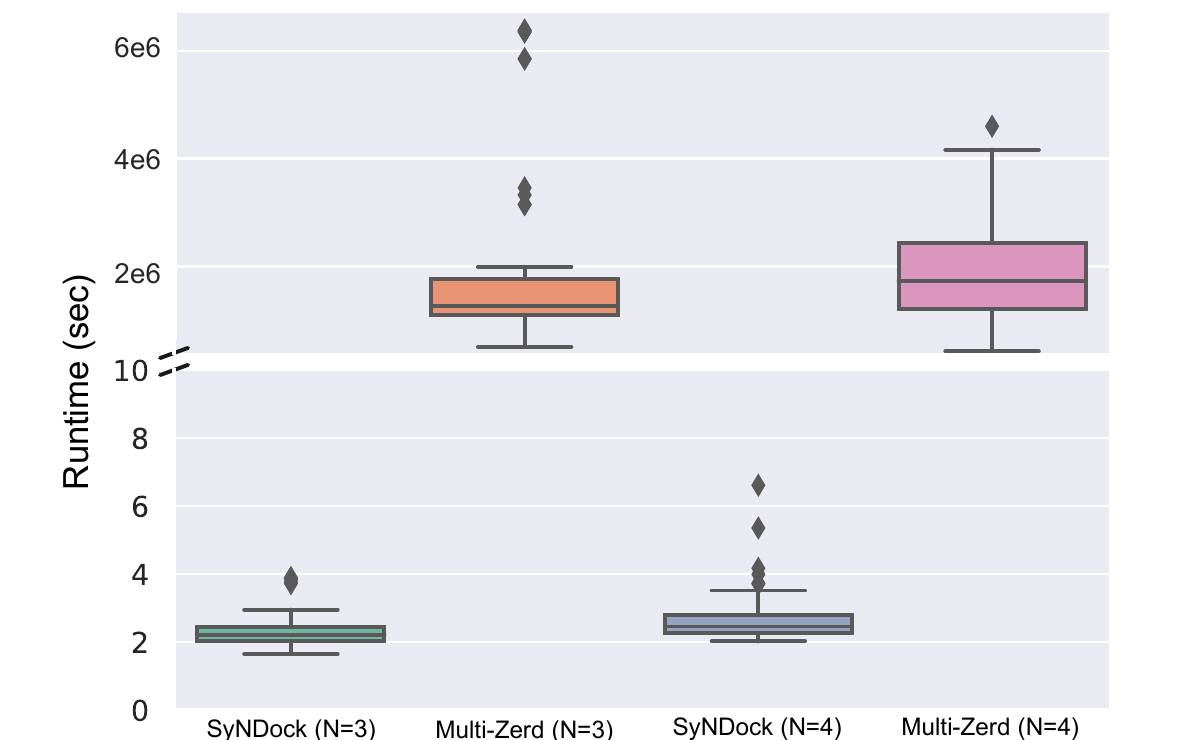}
    \label{fig:exp:speed}
\end{minipage}

\paragraph{Computation Efficiency.}
\cref{fig:exp:speed} compares the inference efficiency of different approaches.
We record the running time of all complex structure prediction algorithms on the trimmer and tetramer test set.
Note that \ours predicts the global transformation in a direct shot, resulting in speeds several orders of magnitude faster than the baseline method.
This is particularly advantageous for intensive screening
applications that need to search over large search spaces, such as drug discovery. 
In addition, as the number of input proteins increases, the superiority of \ours in terms of speed becomes increasingly apparent.

\paragraph{Ablation Study.}
\setlength{\tabcolsep}{1pt}
\renewcommand{\arraystretch}{0.85}
\begin{wraptable}{r}{6.5cm}
    \vspace{-0.4cm}
	\footnotesize
    \centering
       \caption{\textbf{Ablation study} (Trimer test set)}
    \vspace{-0.1cm}
    \begin{tabular}{lcccc} 
    \toprule
    Method & IEMNN & SyNc & C-RMSD & I-RMSD \\
    \hline EquiDock-\textit{Seq} & $\times$ & $\times$ & 25.7 & 26.2 \\
    \hline SyNDock & $\checkmark$ & $\times$ & 25.5 & 20.4 \\
     SyNDock& $\times$ & $\checkmark$ & 22.9 & 19.8 \\
     SyNDock& $\checkmark$ & $\checkmark$ & 22.0 & 18.4 \\
    \bottomrule
    \end{tabular}
    \label{tab:exp:ablation}
\vspace{-0.35cm}
\end{wraptable}
We conduct an ablation study on the trimer test set to assess the significance of the core modules in \ours. As a baseline, we implement a heuristic method called EquiDock-\textit{Seq}, which sequentially docks proteins using a binary docking model (EquiDock).
As shown in \cref{tab:exp:ablation}, EquiDock-Seq fails to effectively address the challenges of multiple protein docking, resulting in C-RMSD and I-RMSD scores of 25.7 and 26.2, respectively.
To evaluate the impact of the learnable synchronization module, we introduce the synchronization module to the pipeline, enabling a learning-centric solution (\ours \textit{wo IEMNN}). This integration significantly improves the performance, achieving C-RMSD and I-RMSD scores of 22.9 and 19.8, respectively.
Next, we investigate the effectiveness of the proposed IEMNN backbone by replacing the backbone of the previous experiments. For the baseline approach (\ours \textit{wo sync}), we observe a slight improvement in I-RMSD to 20.4, while C-RMSD remains unchanged. The best performance is achieved by \ours (standard) with C-RMSD and I-RMSD scores of 22.0 and 18.4, respectively, demonstrates the effectiveness of the IEMNN backbone in aggregating and propagating information across multiple graphs.
Overall, the ablation study highlights the importance and effectiveness of the proposed modules in improving the multimeric protein docking performance.

\paragraph{Visualization}
In \cref{fig:exp:vis}, we show a number of successful examples where \ours has significantly outperformed the baselines on the subset of trimeric proteins.

%% file: src/secs/6-conclusions.tex
\section{Conclusions}

In this paper, we propose a novel multimeric protein docking model, \ours, which allows effective learning to assemble the complex in a globally consistent manner, opening a new perspective on multimeric complex docking.
It enables the automatic assembly of accurate multimeric complexes within a few seconds, the performance of which can be superior or comparable to recent advanced approaches.
Extensive experiments on our curated dataset show that \ours outperforms advanced methods in both accuracy and speed.
Regarding limitations, we acknowledge two main aspects. Firstly, 
limited by the rare number of resolved multibody protein structures, we expanded the number by extracting substructures. However, this  may introduce biases and deviate from the true distribution of protein complexes in nature. This limitation is inherent to the data generation process and can potentially impact the performance and generalizability of the trained models.
Secondly, our approach relies on the rigid assumption, which restricts its applicability to flexible protein docking scenarios.
For future work, we would like to incorporate more domain knowledge and extend the current framework to more applications.

%% file: src/secs/7-app.tex
\section{Notations and Proofs}
\label{supp:sec:notapro}

\subsection{Notations}
We compile a comprehensive list of all the notations utilized in this paper, as shown  in \cref{supp:table:notation}.

\label{notation}
\begin{table}[htp]
    \footnotesize
    \centering
    \caption{Notations.}
    \begin{tabular}{ll}
        \toprule
        Notation & Description \\
        \midrule
        $\mR_{k,l} \in SO(3)$ & the relative rotation matrix between protein $k$ and $l$ \\
        $\mR_k \in SO(3)$ & the absolute rotation matrix of protein $k$ \\
        $\vt_k \in \R^{3}$ & the translation vector of protein $k$ \\
        $\mT_k \in SE(3) \subset \R^{4 \times 4}$ & the transformation matrix of protein $k$ \\
        $\gV_k$ & a set of $n_k$ nodes where each node $v \in \gV_k$ represents an amino acid in protein $k$ \\
        $\gN(i)$ & the set of neighborhood nodes of node $i$  \\
        $\gE_k$ & a set of edges where each  edge $e_{i,j} \in \gE_k$ exists if amino acid $j \in \gN(i)$ in protein $k$ \\
        $\mX_k \in \R^{3 \times n_k}$ & the Cartesian coordinate matrix of the $\alpha$-carbon atoms of a protein $k$ \\
        $\vx_i \in \R^3$ & the Cartesian coordinate vector of a $\alpha$-carbon atom where $\vx_i$ is the $i$-th column of a $\mX$ \\
        $\vh_i \in \R^{f_1}$ & the $f_1$-dimensional feature vector of a node (amino acid) \\
        $\vf_i \in \R^{f_2}$ & the additional $f_2$-dimensional feature vector of a node \\
        $\vf_{j \rightarrow i} \in \R^{f_3}$ & the additional $f_3$-dimensional feature vector of an edge between node $i$ and $j$ \\
        $\gG = (\gV, \gE)$ & a protein graphs where each node represents each amino acid in the protein \\
        $\gS = \{\gG_i, 1 \leq i \leq N\}$ & a set of proteins where each graph $\gG_i = (\gV_i, \gE_i)$ \\
        $\vp_{k,l} \in \R^{3 \times m}$ & pocket points between proteins $k$ and $l$ \\
        $c_{k,l}$ & the estimated confidence between protein $k$ and $l$ \\
        $\mZ_k = \mX_k^{(T)} \in \R^{3 \times n_k}$ & the output coordinate node embeddings of IEMMN for protein $k$ \\
        $\mH_k \in \R^{d \times n_k}$ & the node embeddings of protein $k$ \\
        $u$ &the number of total iterations \\
        \bottomrule
    \end{tabular}
    \label{supp:table:notation}
\end{table}

\subsection{Proofs}
In this part, we present the proof of the theorem proposed in the main paper.

\subsubsection{Proof of \cref{thrm:prop1}}
\begin{proof}
    Assume that $\vh^{(0)}$ is invariant to any orthogonal rotation and translation. By \cref{eq:eq2,eq:eq4}, $\boldsymbol{\mu}_{j \rightarrow i}, \boldsymbol{\mu}$ are obviously invariant.
    
    Note that for any orthogonal rotation $\mQ \in \R^{3 \times 3}$,
    \begin{align*}
        \|\mQ\vx_i^{(t)} - \mQ\vx_j^{(t)}\| = {} & (\vx_i^{(t)} - \vx_j^{(t)})^\top \mQ^\top \mQ(\vx_i^{(t)} - \vx_j^{(t)}) \\
        = {} & (\vx_i^{(t)} - \vx_j^{(t)})^\top \mI(\vx_i^{(t)} - \vx_j^{(t)}) \\
        = {} & \|\vx_i^{(t)} - \vx_j^{(t)}\|^2,
    \end{align*}
    and any translation $\vg \in \R^3$,
    \begin{equation*}
        \|\vx_i^{(t)} + \vg - (\vx_j^{(t)} + \vg)\| = \|\vx_i^{(t)} - \vx_j^{(t)}\|,
    \end{equation*}
    
    Then by \cref{eq:eq1,eq:eq3}, we also have $\vm_{j \rightarrow i}, \vm_i$ are also invariant.
    Therefore, by \cref{eq:eq6} we obtain $\vh_i^{(t+1)}$ is also invariant, which complete the proof.
\end{proof}

\subsubsection{Proof of \cref{thrm:prop2}}

\begin{proof}
    By \cref{thrm:prop1}, we know that $\vm_i$ is invariant to any orthogonal rotation. Then, for any orthogonal rotation $\mQ \in \R^{3 \times 3}$,
    \begin{align*}
        {} & \eta \mQ\vx_i^{(0)} + (1-\eta)\mQ\vx_i^{(t)} + \sum_{j \in \gN(i)}\left(\mQ\vx_i^{(t)} - \mQ\vx_j^{(t)}\right) \phi^{x}\left(\vm_{j \rightarrow i}\right) \\
        = {} & \eta \mQ\vx_i^{(0)} + (1-\eta)\mQ\vx_i^{(t)} + \sum_{j \in \mathcal{N}(i)}\mQ\left(\vx_i^{(l)} -\vx_j^{(t)}\right) \phi^{x}\left(\vm_{j \rightarrow i}\right) \\
        = {} & \eta \mQ\vx_i^{(0)} + (1-\eta)\mQ\vx_i^{(t)} + \mQ\left(\sum_{j \in \mathcal{N}(i)}\left(\vx_i^{(l)} -\vx_j^{(t)}\right) \phi^{x}\left(\vm_{j \rightarrow i}\right)\right) \\
        = {} & \mQ\left(\eta \vx_i^{(0)}+(1-\eta) \vx_i^{(t)} + \sum_{j \in \gN(i)}\left(\vx_i^{(l)}-\vx_j^{(t)}\right) \phi^{x}\left(\vm_{j \rightarrow i}\right)\right) \\
        = {} & \mQ\vx_i^{(t+1)}.
    \end{align*}
\end{proof}

\subsubsection{Proof of \cref{thrm:prop3}}
\begin{proof}
    By \cref{thrm:prop1}, we know that $\vm_i$ is invariant to any translation. Then, for any translation $\vg \in \R^3$,
    \begin{align*}
        {} & \eta(\vx_i^{(0)} + \vg) + (1 - \eta)(\vx_i^{(t)} + \vg) + \sum_{j \in \gN(i)}\left((\vx_i^{(t)} + \vg) - (\vx_j^{(t)} + \vg)\right) \phi^{x}\left(\vm_{j \rightarrow i}\right) \\
        = {} & \eta\vx_i^{(0)} + \eta\vg + (1 - \eta) \vx_i^{(t)} + (1 - \eta)\vg + \sum_{j \in \gN(i)}\left(\vx_i^{(t)} -\vx_j^{(t)}\right) \phi^{x}\left(\vm_{j \rightarrow i}\right) \\
        = {} & \eta\vx_i^{(0)} + (1 - \eta) \vx_i^{(t)} + \sum_{j \in \gN(i)}\left(\vx_i^{(t)} -\vx_j^{(t)}\right) \phi^{x}\left(\vm_{j \rightarrow i}\right) + \vg \\
        = {} & \vx_i^{(t+1)} + \vg.
    \end{align*}
\end{proof}

\subsubsection{Proof of \cref{thrm:prop4}}
\begin{proof}
    For any permutation $\sigma$ on the order of $\mX$,
    denoted by $\sigma(i)$ as the permuted index for $i$. 
    Obviously we have,
    \begin{align*}
        \eta \vx_{\sigma(i)}^{(0)} + (1-\eta) \vx_{\sigma(i)}^{(t)} + \sum_{j \in \mathcal{N}(\sigma(i))}\left(\sigma(\vx_{\sigma(i)}^{(t)})-\sigma(\vx_j^{(t)})\right) \varphi^{x}\left(\mathbf{m}_{j \rightarrow \sigma(i)}\right) %\\
        = {} & \vx_{\sigma(i)}^{(t+1)}
     \end{align*}
\end{proof}

\section{Implementations}
\label{supp:imple}

\subsection{Algorithm}
\label{supp:imple:algo}

As a supplement to method described in the main paper, we provide the implementation
details in \cref{syndock_algo_update} for better clarity. 

\begin{algorithm}[t]
    \KwIn{$\mathcal{G}_k=\left(\mathcal{V}_k, \mathcal{E}_k\right)$}
    \KwOut{$T^{*}(\mR_k, \vt_k)$}
    \SetKwFunction{FMain}{SYNCDock}
    \SetKwProg{Fn}{Function}{:}{}
    \Fn{\FMain{$F$}}{
        Extract multi-graph features using the IEMMN backbone\;
        Compute the pairwise relative transformations $\mT_{k, l}$ and the docking confidence scores $w_{k, l}$\;
        \For{$\text{iter} = 1$ \KwTo $4$}{
            Compute the absolute transformations $\mT_k = (\mR_k, \vt_k)$ for all $1 \leq k \leq M$ using the transformation synchronization module described in \cref{NagetionAlgo}\; 
            Update the graph $\mathcal{G}_k$ based on the current estimate of $\{\mR_k, \vt_k\}$\;
        }
        
        Compute the loss based on the updated transformations and the desired outcome (during training)\;
        
        \textbf{return} the optimized transformations $T^{*}(\mR_k, \vt_k)$\;
    }
    \textbf{End Function}
    \caption{SyNDock}
    \label{syndock_algo_update}
\end{algorithm}

\begin{algorithm}[t]
    \KwIn{$F(w_{k, l}, \mT_{k, l}), \forall (k, l) \in \gE$}
    \KwOut{$T^{*}(\mR_k, \vt_k), \forall k \in \gV$}
    \SetKwFunction{FMain}{SYNC}
    \SetKwProg{Fn}{Function}{:}{}
    \Fn{\FMain{$F$}}{
        Compute the weighted connection Laplacian $\mL$ and vector $\vb$\;
        Calculate the first three eigenvectors $\mU$ of $\mL$\;
        Perform Singular Value Decomposition (SVD) on blocks of $\mU$ to obtain $\{\mR_k, 1 \leq k \leq N\}$\;
        Solve the synchronization problem defined in \cref{sync_t} to obtain $\{\vt_k, 1 \leq k \leq N\}$\;
        \textbf{return} $\mT_k = (\mR_k, \vt_k), 1 \leq k \leq M$\;
    }
    \textbf{End Function}
    \caption{Transformation Synchronization Module}
    \label{NagetionAlgo}
\end{algorithm}

\subsection{Dataset}
The overview of datasets is in \cref{tab:exp:dataset}. DIPS is downloaded from \url{https://github.com/drorlab/DIPS}. The DIPS dataset (Distant Interacting Protein Structures)~\cite{dips} is constructed by mining the Protein Data Bank (PDB) and serves as a comprehensive collection of protein complex structures where individual protein structures are unavailable. The dataset comprises a curated set of 42,462 binary protein interactions.
For this study, we utilized the latest version of the PDB, which provides a total of 282,076 PDB files. To filter the data, we applied the following criteria: a) a buried surface area of more than 500, b) a resolution of NMR structures less than 3.5, c) a sequence identity of less than 30, and d) a complex size exceeding 50.0.
The raw PDB files underwent several preprocessing steps as follows:
\begin{itemize}
    \item Valid residues were identified based on the presence of "Natom," "alphaCatom," and "Catom" attributes.
    \item The coordinates of the "alphaCatom" were extracted to obtain the residue coordinates.
    \item Pairwise distances between residues in the two protein arrays were computed. If the distance between residues was less than 8.0, the residues were marked as pockets. Additionally, the pocket ground truth was obtained by averaging the coordinates of the identified pockets from the two proteins.
\end{itemize}

\begin{figure}[h!]
	\centering
	\includegraphics[width=\linewidth]{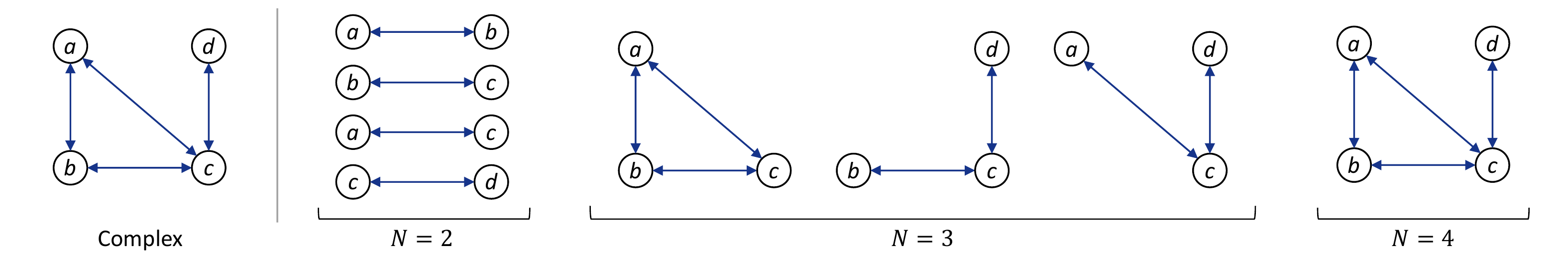}
	\caption{An example of our use of the dataset was obtained by curating from the DIPS dataset: we searched through the proteins to obtain connected subgraphs of different degrees}
	\label{fig:dataset:curation}
\end{figure}

After performing the above steps, we grouped the docked protein chains with the same PDB and performed subgraph mining on them as shown in  \cref{fig:dataset:curation} to obtain the multibody complexes.

\subsection{Data representation}
We follow previous work~\cite{jing2020learning,ganea2022independent} to represent protein as a graph.
We construct the graph using the K-nearest neighbors (KNN) graph construction method.
We adopt the feature construction used in~\cite{ganea2022independent} for feature encoding.